\documentclass[preprint,prc,showpacs,preprintnumbers,superscriptaddress]{revtex4}

\usepackage{dcolumn}
\usepackage{bm}
\usepackage{longtable}
\usepackage{dsfont}

 \usepackage{graphicx,epsfig,latexsym,amssymb}
 \usepackage{multirow,amsmath,array,booktabs,color}
 \usepackage[section]{placeins}

\begin{document}

 \title{Candidate M$\chi$D nucleus $^{106}$Rh in triaxial relativistic
 mean-field approach with time-odd fields }

\author{J. M. Yao }
\address{School of Phyics and State Key Laboratory of Nuclear Physics and
 Technology, Peking University, 100871 Beijing, China}

\author{B. Qi}
\address{School of Phyics and State Key Laboratory of Nuclear Physics and
 Technology, Peking University, 100871 Beijing, China}
\author{S. Q. Zhang }
\address{School of Phyics and State Key Laboratory of Nuclear Physics and
 Technology, Peking University, 100871 Beijing, China}

\author{J. Peng }
\address{Department of Physics, Beijing Normal University, 100875 Beijing, China}

 \author{S. Y. Wang }
 \address{Department of Space Science and Applied Physics, Shandong University
 at Weihai, 264209, Weihai, China}

 \author{J. Meng}
 \email{mengj@pku.edu.cn}
 \address{School of Phyics and State Key Laboratory of Nuclear Physics and
              Technology, Peking University, 100871 Beijing, China}
 \address{Institute of Theoretical Physics, Chinese Academy of
              Sciences, Beijing, China}
 \address{Center of Theoretical Nuclear Physics, National Laboratory of Heavy
      Ion Accelerator, 730000 Lanzhou, China}
 \address{ Department of Physics, University of Stellenbosch, Stellenbosch, South
 Africa}

\date{\today}

\begin{abstract}

 The configuration-fixed constrained triaxial relativistic mean-field approach
 is extended by including time-odd fields and applied to study the
 candidate multiple chiral doublets (M$\chi$D) nucleus $^{106}$Rh.
 The energy contribution from time-odd fields and microscopical evaluation of center-of-mass
 correction as well as the modification of triaxial deformation
 parameters $\beta,\gamma$ due to the time-odd fields are investigated.
 The contributions of the time-odd fields to the total energy
 are 0.1-0.3 MeV and they
 modify slightly the $\beta,\gamma$ values. However,
 the previously predicted multiple chiral
 doublets still exist.

\end{abstract}
 \maketitle

Since the prediction of existence of chirality in atomic nuclei in
1997~\cite{Frauendorf97} and later experimental observation of
chiral doublet bands in 2001~\cite{Starosta01}, nuclear chirality
has become one of the most interesting subjects in nuclear physics.
Hitherto, extensive studies have been performed to understand the
phenomena and explore their possible existence in $A \sim 100, 130$
and $190$ mass
regions~\cite{Tonev06,Grodner06,Mukhopadhyay07,Joshi07,Meng08}.

 On the theoretical side, chiral doublet bands were first predicted by
 the particle-rotor model (PRM) and tilted axis cranking (TAC) model for
 triaxially deformed nuclei~\cite{Frauendorf97}.
 Later on, numerous efforts have been devoted to the development
 of TAC methods~\cite{Dimitrov00,Olbratowski04,Olbratowskiprc,Peng08}
 and PRM models~\cite{Peng03a,Koike04,Wang07,Zhang07} to describe chiral rotation
 in atomic nuclei. It is shown that triaxial deformation and high-$j$ valence
 particles and valence holes are essential for the formation of chirality in nuclei.
 Therefore, it will be very interesting to search for nuclei with these characters
 within the state-of-the-art nuclear structure models.

Relativistic mean-field (RMF)
theory~\cite{Serot86,Reinhard89,Ring96,Vretenar05,Meng06ppnp}, which
relies on basic ideas of effective field theory and of density
functional theory has achieved great success in describing many
nuclear phenomena for both stable and exotic nuclei over the entire
nuclear chart. It thus provides us a microscopic way to study
nuclear structure properties including the energy and deformation
for not only the ground state but also the excited state for given
valence nucleon configuration. In Ref.~\cite{Meng06}, a
configuration-fixed constrained triaxial RMF approach was developed
and applied to study the nuclear potential energy surface (PES). An
interesting phenomenon -- the existence of multiple chiral doublets
(M$\chi$D), i.e., more than one pair of chiral doublet bands in one
single nucleus, has been suggested in $^{106}$Rh and other odd-odd
Rhodium isotopes~\cite{Peng08}. These predictions are based on the
triaxial deformations of local minima and the corresponding proton
and neutron configurations. In these studies, the time-reversal
invariance was assumed from the beginning, namely, the time-odd
fields were neglected.

Actually, the unpaired valence neutron and proton will generate
nucleon currents and break the time-reversal invariance in nuclear
state. Such effects have been found to be of great importance to
reproduce the nuclear magnetic moment~\cite{Hofmann88}, inertia of
moment~\cite{Konig93} as well as $M1$ transition rates in magnetic
rotation nuclei~\cite{Peng08-2}. Therefore one has to examine the
existence of M$\chi$D in odd-odd Rhodium isotopes with the presence
of time-odd fields.

In this work, the configuration-fixed constrained triaxial RMF
approach will be extended by including time-odd fields, which is
more suitable to study the triaxial structure properties of odd-mass
and odd-odd nuclei. Taking $^{106}$Rh as an example, the effect of
time-odd fields on the total energy, triaxial deformations
$\beta,\gamma$ as well as configuration will be examined.

 The detailed description of configuration-fixed constrained triaxial RMF approach
 with nucleon-nucleon interacting via meson exchange can be found in Ref.~\cite{Meng06}
 and references therein. Only a brief outline, in particular with the presence of time-odd
 fields, will be given here.

 The starting point of the RMF theory
 is the standard effective Lagrangian density constructed with the degrees
 of freedom associated with nucleon field~($\psi$), two isoscalar meson fields
 ~($\sigma$ and $\omega_\mu$), isovector meson field~($\vec\rho_\mu$)
 and photon field~($A_\mu$). Under ``mean-field" and ``no-sea" approximations, one can
 derive the corresponding energy density functional, from which one finds immediately
 the equation of motion for a single-nucleon orbit $\psi_i(\bm{r})$ with the help of variational
 principle,
  \begin{equation}
  \label{DiracEq}
  \{\mathbf{\alpha}\cdot[\bm{p}- \bm{V} (\bm{r})]
  +\beta m^*(\bm{r})+V_0(\bm{r})\}\psi_i(\bm{r})
  =\epsilon_i\psi_i(\bm{r}),
 \end{equation}
 where $m^*(\bm{r})$ is defined as $m^*(\bm{r})\equiv m+g_\sigma\sigma(\bm{r})$,
 with $m$ referring to the mass of bare nucleon. The repulsive vector potential $V_0(\bm{r})$,
 i.e., the time-like component of vector potential reads,
 \begin{equation}
  \label{vecpot}
  V_0(\bm{r})=g_\omega\omega_0(\bm{r})+g_\rho\tau_3\rho_0(\bm{r})
  +e\frac{1-\tau_3}{2}A_0(\bm{r}),
 \end{equation}
 where  $g_i(i=\sigma,\omega,\rho)$
 are the coupling strengthes of nucleon with mesons.
 The time-odd fields $\bm{V}(\bm{r})$ are naturally given by the space-like components
 of vector fields,
 \begin{equation}
  \label{magpot}
   \bm{V}(\bm{r})=g_\omega\bm{\omega}(\bm{r})
                          +g_\rho\tau_3\bm{\rho}(\bm{r})
                          +e\frac{1-\tau_3}{2}\bm{A}(\bm{r}).
 \end{equation}

 The non-vanishing time-odd fields in Eq.(\ref{magpot}) give rise to splitting
 between pairwise time-reversal states $\psi_{\overline i}$ and
 $\psi_{\underline i}(\equiv\hat T\psi_{\overline i})$, where $\hat T$
 is the time-reversal operator. Each Dirac spinor $\psi_i(\bm{r})$ is expanded
 in terms of a set of three-dimensional harmonic oscillator (HO) basis in Cartesian
 coordinates with 12 major shells. The meson fields which provide the nuclear mean-field
 potentials are expanded in terms of the same HO basis
 as those of Dirac spinor but with 10 major shells.
 The pairing correlations are greatly quenched by the unpaired valence neutron
 and proton in $^{106}$Rh and thus neglected. More details about the
 solution of Dirac equation (\ref{DiracEq}) with time-odd fields can be found in
 Ref.~\cite{Yao06}.

 A configuration-fixed quadrupole moment constraint calculation
 through $\beta^2$ was carried out to obtain the PES
 for each configuration, where $\beta=\displaystyle\frac{4\pi}{3A R^2_0} \sqrt{q^2_{20}+2q^2_{22}}$
 and $\gamma=\tan^{-1}(\sqrt{2}\displaystyle\frac{q_{22}}{q_{20}})$
 with $ q_{20}=\displaystyle\sqrt{\frac{5}{16\pi}}\langle 2z^2-x^2-y^2\rangle$
 and $q_{22} = \displaystyle\sqrt{\frac{15}{32\pi}}\langle x^2-y^2\rangle$.
 The same configuration is guaranteed during the procedure of constraint calculation
 with the help of ``parallel-transport"~\cite{Bengtsson89},
 which enables one to decompose the whole PES into several parts
 characterized by the quantum numbers of corresponding configurations.%

 \begin{figure}[h!]
 \centering
 \includegraphics[width=6cm]{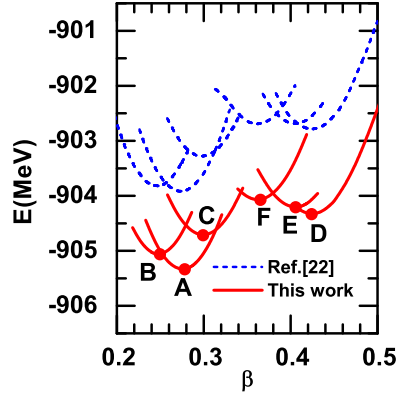}
  \caption{(Color online) The potential energy surfaces as functions of deformation $\beta$ in
    configuration-fixed constrained time-odd triaxial RMF
    calculations with PK1 set (solid line). The minima in the energy
    surfaces of each configuration are labeled with A, B, C, D, E, and F
    respectively according to their energies.
    The results by triaxial RMF calculation without time-odd fields
    (dashed line) are taken from Ref.~\cite{Peng08}.}
 \label{fig1}
 \end{figure}

 In Fig.~\ref{fig1}, the energies are given as functions of deformation $\beta$
 in configuration-fixed constrained time-odd triaxial RMF calculations with
 PK1 set~\cite{Long04} for $^{106}$Rh. The minima in the energy
 surfaces of each configuration are labeled with A,
 B, C, D, E, and F respectively. The PES plotted with dashed line in Fig.~\ref{fig1}
 are obtained by triaxial RMF calculation without time-odd fields.
 Furthermore the center-of-mass (c.m.) correction energy is estimated
 phenomenologically with $E^{\rm phe.}_{\rm c.m.}=-\dfrac{3}{4}\times41A^{-1/3}$,
 which remains to be a constant
 for all configurations. Here the c.m. correction energy is
 evaluated microscopically by projection-after-variation in the first-order
 approximation, i.e.,
 \begin{equation}
  E^{\rm mic.}_{\rm c.m.}
  =-\dfrac{1}{2mA}\langle \bm{P}^2_{\rm c.m.}\rangle,
 \end{equation}
 where $\bm{P}_{\rm c.m.}=\sum_i^A \bm{p}_i$ and $A$ is the mass
 number. It is found that the time-odd fields and microscopic c.m. correction
 do not change significantly the topological structure of the whole PES but lower it down
 about 1.5 MeV. As a result, the energy of ground state is modified
 from -903.92 MeV to -905.33 MeV, which is much closer to the experimental
 data -906.72 MeV~\cite{Audi03}.
\begin{figure}[h!]
 \centering
 \includegraphics[width=6cm]{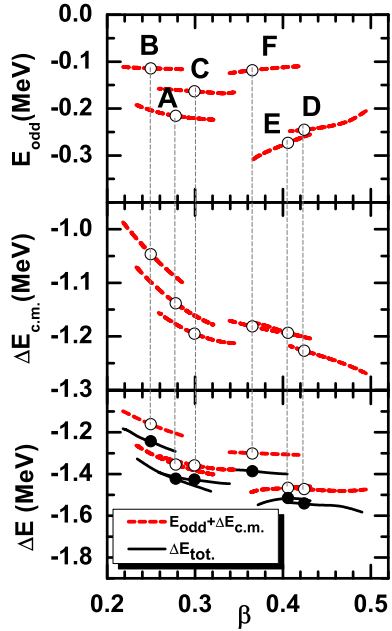}
  \caption{(Color online) The energy from time-odd fields $E_{\rm odd}$(upper
  panel), the energy difference between microscopic
  and phenomenological c.m. correction $\Delta E_{\rm c.m.}$
  (middle panel), the summation of $E_{\rm odd}$ and $\Delta E_{\rm c.m.}$ (dashed line)
  and the energy difference $\Delta E_{\rm tot.}$ (solid line) between the corresponding
  potential energy
  surfaces in Fig.~\ref{fig1} (lower panel) for different configurations
  as functions of deformation parameter $\beta$.}
 \label{fig2}
 \end{figure}

 In Fig.~\ref{fig2}, we plot the energy contribution from the
 time-odd fields $E_{\rm odd}$
 $(\equiv-\dfrac{g_\omega}{2}\int d^3\bm{r}\bm{\omega}(\bm{r})\cdot
 \mathbf{j}_N(\bm{r})$, with the nucleon current given by
 $\mathbf{j}_N=\sum_i\psi^\dagger_i\bm{\alpha}\psi_i$),
 the c.m. correction energy difference $\Delta E_{\rm c.m.}=E^{\rm mic.}_{\rm c.m.}-E^{\rm phe.}_{\rm c.m.}$ and
 total energy difference $\Delta E_{\rm tot}$ between the present calculation
 and those in Ref.~\cite{Peng08} as functions of
 deformation parameter $\beta$. The dashed line in lower panel of Fig.~\ref{fig2} denotes the
 summation of $E_{\rm odd}$ and $\Delta E_{\rm c.m.}$.
 All $E_{\rm odd}$, $\Delta E_{\rm c.m.}$ and $\Delta E_{\rm tot}$
 change moderately as functions of deformation parameter $\beta$ for
 given configuration. Furthermore, the
 main contribution to $\Delta E_{\rm tot}$, i.e., the shift of whole PES in
 Fig.~\ref{fig1} is due to $\Delta E_{\rm c.m.}$. The time-odd
 fields make the nucleus more bound and their contributions to energy range
 around 0.1-0.3 MeV. From the lower panel in Fig.~\ref{fig2}, it tells us
 that the time-odd fields will modify the
 time-even mean-fields and lead to $\sim 0.1 $MeV contribution to the
 total energy for all the configurations.

 \begin{figure}[h!]
 \centering
 \includegraphics[width=8cm]{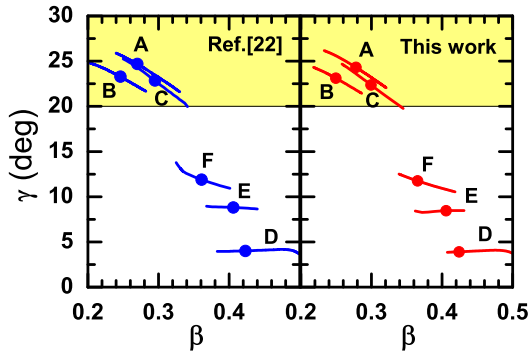}
  \caption{(Color online) The triaxial deformation parameters $\gamma$ as functions of $\beta$ in
    configuration-fixed constrained triaxial RMF calculations for
    $^{106}$Rh without (left) or with (right)  the time-odd fields.
    The results in the left panel are taken from Ref.~\cite{Peng08}.}
 \label{fig3}
 \end{figure}

In Fig.~\ref{fig3}, we plot the triaxial deformation parameters
$\gamma$ as functions of $\beta$ in configuration-fixed constrained
triaxial RMF calculations for $^{106}$Rh without (left panel) or
with (right panel) the time-odd fields. The shaded area represents
favorable triaxial deformation for chirality. It shows that triaxial
deformation parameters $\beta$ and $\gamma$ are not sensitive to the
time-odd fields. In both cases, the valence nucleon configurations
A, B and C have the favorable triaxial deformation for chirality.

In order to label the configuration A, B and C, the main spherical
component for the wave function of the valence nucleon has been
obtained by expanding the Dirac spinor in terms of spherical HO
basis with the quantum number $\vert nljm\rangle$. It is found that
the influence of the time-odd fields for the composition of the
Dirac spinor is negligible.

 \begin{table*}[]
   \centering
   \tabcolsep=4pt
   \caption{The total energies $E_{\rm tot.}$, center-of-mass correction
   energy $E_{\rm c.m.}$, energy contribution from the
   time-odd fields $E_{\rm odd}$, triaxial deformation parameters
   $\beta,\gamma$ as well as their corresponding valence nucleon
   configurations for A-F in the configuration-fixed constrained triaxial RMF calculations
   with (without) time-odd fields. The values in parentheses
   are taken from Ref.~\cite{Peng08}.}
   \begin{tabular}{ |c|cccccc|}
   \toprule
   State &configuration                            &$E_{\rm tot.}$(MeV) & $E_{\rm c.m.}$(MeV)  & $E_{\rm odd}$(MeV)           &  $\beta $ & $\gamma $                           \\
   \hline
   A     &$\nu2d^{1}_{5/2}\otimes\pi1g^{-3}_{9/2}$ & -905.33 (-903.92)  &  -7.64 (-6.50)  &  -0.22                  & 0.28 (0.27)  & 24.3$^\circ$ (24.7$^\circ$)    \\
   B     &$\nu1h^{1}_{11/2}\otimes\pi1g^{-3}_{9/2}$& -905.06 (-903.82)  &  -7.55 (-6.50)  &  -0.11                  & 0.25 (0.25)  & 23.1$^\circ$ (23.3$^\circ$)     \\
   C     &$\nu1h^{3}_{11/2}\otimes\pi1g^{-3}_{9/2}$& -904.72 (-903.28)  &  -7.69 (-6.50)  &  -0.16                  & 0.30 (0.30)  & 22.4$^\circ$ (22.9$^\circ$)     \\
   D     &$\nu1h^{5}_{11/2}\otimes\pi2p^{-1}_{3/2}$& -904.33 (-902.79)  &  -7.73 (-6.50)  &  -0.25                  & 0.42 (0.42)  & 3.9$^\circ$  (4.0$^\circ$)     \\
   E     &$\nu1g^{-1}_{7/2}\otimes\pi2p^{-1}_{3/2}$& -904.21 (-902.68)  &  -7.69 (-6.50)  &  -0.27                  & 0.41 (0.41)  & 8.5$^\circ$  (8.8$^\circ$)     \\
   F     &$\nu1g^{-1}_{7/2}\otimes\pi1g^{1}_{7/2}$ & -904.07 (-902.69)  &  -7.68 (-6.50)  &  -0.12                  & 0.37 (0.36)  & 11.8$^\circ$ (11.9$^\circ$)     \\
   \hline\hline
 \end{tabular}
 \label{tab1}
 \end{table*}

The total energies $E_{\rm tot.}$, center-of-mass correction energy
$E_{\rm c.m.}$, energy contribution from the time-odd fields $E_{\rm
odd}$, triaxial deformation parameters $\beta,\gamma$ as well as
their corresponding valence nucleon configurations for A-F in the
configuration-fixed constrained triaxial RMF calculations with
(without) time-odd fields are presented in Tab.~\ref{tab1}. It shows
that the time-odd fields may reduce the triaxial deformation
parameter $\gamma$ by 0.5$^\circ$. Using the structure information
for the configuration $\nu h^{1}_{11/2}\otimes\pi g^{-3}_{9/2}$ as
inputs in a triaxial rotor coupled with quasiparticles
model~\cite{Wang07,Zhang07}, the energy spectra and the
electromagnetic transition ratios for a pair of negative-parity
doublet bands in $^{106}$Rh are well-reproduced in
Ref.~\cite{Wang08}.

In summary, the configuration-fixed constrained triaxial
relativistic mean-field approach has been extended by including
time-odd fields and applied to study the  candidate M$\chi$D nucleus
$^{106}$Rh. The energy contribution from time-odd fields and
center-of-mass correction has been studied in detail. It has been
found that the time-odd fields contribute 0.1-0.3 MeV to the total
energy and slightly modify the triaxial deformation parameters
$\beta,\gamma$. It confirms the previous prediction of possible
existence of M$\chi$D in configuration-fixed triaxial RMF approach
without time-odd fields. As one pair of doublet bands with $\nu
h^{1}_{11/2}\otimes\pi g^{-3}_{9/2}$ configuration has been observed
experimentally, it will be very interesting to search for the
candidate chiral doublet bands with configuration $\nu
h^{3}_{11/2}\otimes\pi g^{-3}_{9/2}$ and verify the prediction of
M$\chi$D.

\vspace{2em}This work is partly supported by the National Natural
 Science Foundation of China under Grant No. 10775004, 10705004,
 10875074, and 10505002
 and the Major State Basic Research Development Program
 Under Contract Number 2007CB815000.


\end{document}